\documentclass[12pt]{article}
\pdfoutput=1
\usepackage{amssymb,amsmath,graphicx,color}
\begin{document}
\title{\textbf{Approaching a strong fourth family}} 
\author{B.~Holdom%
\thanks{bob.holdom@utoronto.ca}\\
\emph{\small Department of Physics, University of Toronto}\\[-1ex]
\emph{\small Toronto ON Canada M5S1A7}}
\date{}
\maketitle
\begin{abstract}
 A heavy fourth family is an example of new physics which is well defined and familiar in some respects, but which nevertheless has radical implications. In particular it eliminates a light Higgs description of electroweak symmetry breaking. We discuss an early signal for heavy quarks at the LHC in the form of an excess of ``$W$-jets'', and as well show how $W$-jets may be useful in the reconstruction of the heavy quark masses. We argue that fourth family quarks can be distinguished from vector-like quarks of a similar mass at roughly the same time that a same sign lepton signal becomes visible. Given the large mass of the fourth neutrino we describe how a picture for neutrino mass emerges in the absence of right-handed neutrinos, and how it suggests the existence of a remnant flavor gauge symmetry. Based on talk given at ``Second Workshop on Beyond 3 Generation Standard Model --- New Fermions at the Crossroads of Tevatron and LHC'', January 2010, Taipei Taiwan.
\end{abstract}
\section{Motivation}
A recent result from the Tevatron places a lower bound of 338 GeV on the mass of the $b'$ quark of a fourth family  \cite{a0}. This lower bound is narrowing the allowed mass window for fourth family quarks, since the upper bound allowed by partial wave unitarity is about 600 GeV \cite{a4}. Nevertheless, depending on whether the masses are at the low or high end of this range a fourth family could still be characterized as ``light'' or ``heavy''. The implications are profoundly different. The response to the discovery of a light fourth family may be ``Who ordered \textit{that}?'', while the discovery of a heavy family could cause a realization as in  ``So \textit{that} is the way nature works!''

A light fourth family is introduced into the standard model in the hope that it can co-exist with the Higgs boson. The new family significantly modifies the running of the quartic Higgs coupling as reflected by the two new terms in $\mu d\lambda/d\mu\propto \lambda y_{q'}^2-y_{q'}^4+...$ involving the new large Yukawa couplings. These terms must be carefully balanced to keep $\lambda(\mu)$ finite and positive at some higher scale $\mu=\Lambda$, and this translates into a smaller allowed range of the Higgs mass \cite{a2a}. The Yukawa couplings $y_{q'}(\mu)$ also tend to quickly run into trouble at higher scales. Perhaps these issues can be side-stepped if the Yukawa and Higgs couplings instead approach a nontrivial ultraviolet fixed point \cite{a5}.

More troublesome is the direct contribution to the Higgs mass from the loop of fourth family fermions. If we are to take this seriously then there must be some new physics which acts to cut off this quadratically divergent contribution. Assuming this effective cutoff is somewhat above the heavy quark masses, then these quarks act to shift the Higgs mass by
\begin{equation}
\delta m_h^2\approx \left[\frac{m_{q'}}{\mbox{400 GeV}}\right]^2\Lambda^2.
\end{equation}
Thus as the heavy quark masses approach 400 GeV, the Higgs mass in the absence of fine tuning is pushed up to the cutoff. When the Higgs mass is pushed up to the scale where new physics must enter, then the original Higgs description is no longer useful.

When a fourth family becomes clearly ``heavy'' then it clearly cannot co-exist with a light Higgs.\footnote{The cross-over mass which separates ``light'' from ``heavy'' remains quite uncertain.} When $m_{q'}\approx$ 600 GeV then the Goldstone bosons of electroweak symmetry breaking couple maximally strongly to the $t',b'$. The self-interactions of the Goldstone bosons will also be strong and so strong interactions rather than a weak Higgs exchange will unitarize $WW$ scattering. A Higgs not only loses meaning but it is also no longer needed, since the strong interactions can cause dynamical symmetry breaking and thus yield the Goldstone bosons directly. Following the cue of QCD, we can expect that a bilinear fermion condensate takes the place of the Higgs vacuum expectation value.\footnote{For similar discussion along these lines and for references to earlier work see \cite{a1}.}

We also need an alternative to Yukawa couplings for the generation of quark and lepton masses. In the presence of electroweak symmetry breaking fermion condensates of size $\Lambda_{ew}^3$, appropriate 4-fermion operators can feed mass down to other quarks and leptons. The origin of these 4-fermion operators imply the existence of new scales of flavor physics above the TeV scale. If such a scale is $\Lambda_{fl}$ then the resulting quark or lepton mass is of order $\Lambda_{ew}^3/\Lambda_{fl}^2$. What then causes fermions to develop the needed condensates? A new confining gauge interaction, technicolor, is one possibility, but another is to have new strong \textit{broken} gauge interactions. It is for this latter possibility that we can identify the fermions developing condensates with the fourth family, since by definition a sequential fourth family does not feel any new unbroken gauge interaction.

The idea that new broken gauge interactions are involved with electroweak symmetry breaking is also economical. We have just mentioned that new flavor interactions well above the TeV scale are responsible for feeding mass to light quarks and leptons. The gauge symmetry broken near a TeV may then just be a remnant of this larger flavor gauge symmetry, where the latter only partially breaks at the higher scale. The remnant flavor interactions must yield effective 4-fermion operators that are strong enough to produce the electroweak scale condensates. Thus broken flavor gauge interactions may hold the key to both electroweak symmetry breaking and quark and lepton masses.

We return to the point that a heavy fourth family is incompatible with a light Higgs. This provides strong motivation to search for a fourth family, given that the experimental signatures of a fourth family at the LHC are more accessible than those of a light Higgs. It then becomes intriguing to wonder whether it is possible to rule out a light Higgs before the search for it gets seriously under way. The answer to this question involves another question. How easy is it to tell a sequential quark of a fourth family from a ``vector quark'' (vector-like in the sense that its mass preserves electroweak symmetry)? Unlike sequential quarks, vector quarks are compatible with and are sometimes even motivated by a light Higgs. Vector quarks have unconstrained masses and they have decay modes involving the $Z$ as well as the $W$. We shall consider a method to distinguish vector quarks from sequential quarks in the next section.

\section{Searches}
We first would like to consider what the early indications of a sequential fourth family at the LHC may look like. We shall assume that the dominant decay modes are 
\begin{equation}
t'\rightarrow W^+b\quad \mbox{and}\quad b'\rightarrow W^-t
.\end{equation}
We shall also assume that a $b$-tagger is not yet operational in the early searches. We can consider signatures that are indirect in the sense that they do not involve a full mass reconstruction of the heavy quark. A well studied method of this type is the search for same sign leptons from $b'\overline{b'}\rightarrow W^+W^-W^+W^-b\overline{b}$, where the backgrounds are believed to be small \cite{a6,a0}. But same sign leptons could also be a signal of other types of new physics, and so this search needs to be supplemented by other approaches.

A strategy to be explored here is to focus on the excess of boosted and isolated $W$'s, that arise from both $b'\overline{b'}\rightarrow W^+W^-W^+W^-b\overline{b}$ and $t'\overline{t'}\rightarrow W^+W^-b\overline{b}$. The hadronic decay of such $W$'s give rise to $W$-jets, and they have been studied in the context of the mass reconstruction of the heavy quarks \cite{a2,a21,a3}. Here we shall study the excess of $W$-jets directly by looking for a peak in the jet invariant mass distribution. To suppress backgrounds one of the other $W$'s is required to decay leptonically. In each event we shall consider the jet with the largest jet mass to construct the jet mass distribution.

Event generation and simulation details are as follows. Signal and background events are generated by Madgraph \cite{C} and Alpgen \cite{B} respectively, along with Pythia \cite{A} using tune 108 (D6) with CTEQ6L1. The primary backgrounds of interest are $t\overline{t}+\rm{jets}$ and $W+\rm{jets}$, and MLM jet matching is used with $p_{T{\rm min}}=100$ and 150 GeV respectively. Detector simulation is performed with PGS \cite{D} modified to use the anti-$k_T$ jet algorithm and using CMS settings.\footnote{Results using ATLAS settings are similar and can be found at \cite{E}.} We present results at $\sqrt{s}= 10$ TeV for $m_{q'}=600$ and at $\sqrt{s}= 7$ TeV for $m_{q'}=450$ GeV. Results are always normalized to a luminosity of 1 fb$^{-1}$.

We use a $K$ factor of 1.5 for both the pair production of heavy quarks and the $t\overline{t}+\rm{jets}$ background, as indicated by the NLO and NLL cross sections in \cite{a1a}. These cross section results do not directly apply to the relevant $t\overline{t}+\rm{jets}$ background, where $H_T$ is large compared to the $t\overline{t}$ threshold, but the study in \cite{a2} comparing MC@NLO to Alpgen suggests that $K=1.5$ is reasonable. Our $W+\rm{jets}$ background is dominated by 2 or more jets; for $W+\mbox{2 jets}$ a $K$ factor somewhat less than unity was found \cite{a1b}, and for $W+\mbox{3 jets}$ the $K$ factor may be even less \cite{a1c}. Nevertheless we set $K=1$ for this background.

For event selection we require three or more jets with $p_T>100$ GeV and $|\eta|<2.5$, as well as an $L$ defined as an isolated lepton with $p_T>15$ GeV \textbf{or} missing $E_T>200$ GeV. We focus on the jet with largest invariant mass in each selected event. This jet is required to be isolated, separated by $\Delta R>1$ from the other $p_T>100$ GeV jets and $L$, and if so its mass is used to form the histogram.

There are two quantities that need to be adjusted to enhance the signal to background ratio: the value of $H_{T{\rm min}}$ for the $H_T$ cut and the jet resolution parameter $R$. The optimal $H_{T{\rm min}}$ is about $2m_{q'}$. $R$ should be large enough to make it likely that a single jet can capture the hadronic decay products of a $W$ produced in a heavy quark decay. Thus the optimal $R$ also depends on $m_{q'}$, becoming smaller for larger $m_{q'}$. Therefore the observed effect that the variation of $H_{T{\rm min}}$ and $R$ has on the signal can provide information about $m_{q'}$.

The signal events are shown in Fig.~(1) where it is seen that the $t'\overline{t'}$ and $b'\overline{b'}$ production contribute about equally to the strong peak at the $W$ mass. $b'\overline{b'}$ production has a larger high jet mass tail due to the presence of the $t$ in the final state. Fig.~(2) shows the combined signal and background for $H_{T{\rm min}}=2m_{q'}=1200$ GeV and $R=0.7$. Signal to background is degraded for a smaller $H_{T{\rm min}}$, as seen in Fig.~(3) where $H_{T{\rm min}}=1000$ GeV. The procedure of optimizing the results as a function of $H_{T{\rm min}}$ will thus help to determine $m_{q'}$. In Fig.~(4) $R$ is increased to 0.8, and this shows the sensitivity of the shape of the jet mass distribution to this quantity. Increasing $R$ moves the distribution towards higher values as expected, but the signal remains strong.
\begin{center}\includegraphics[scale=0.4]{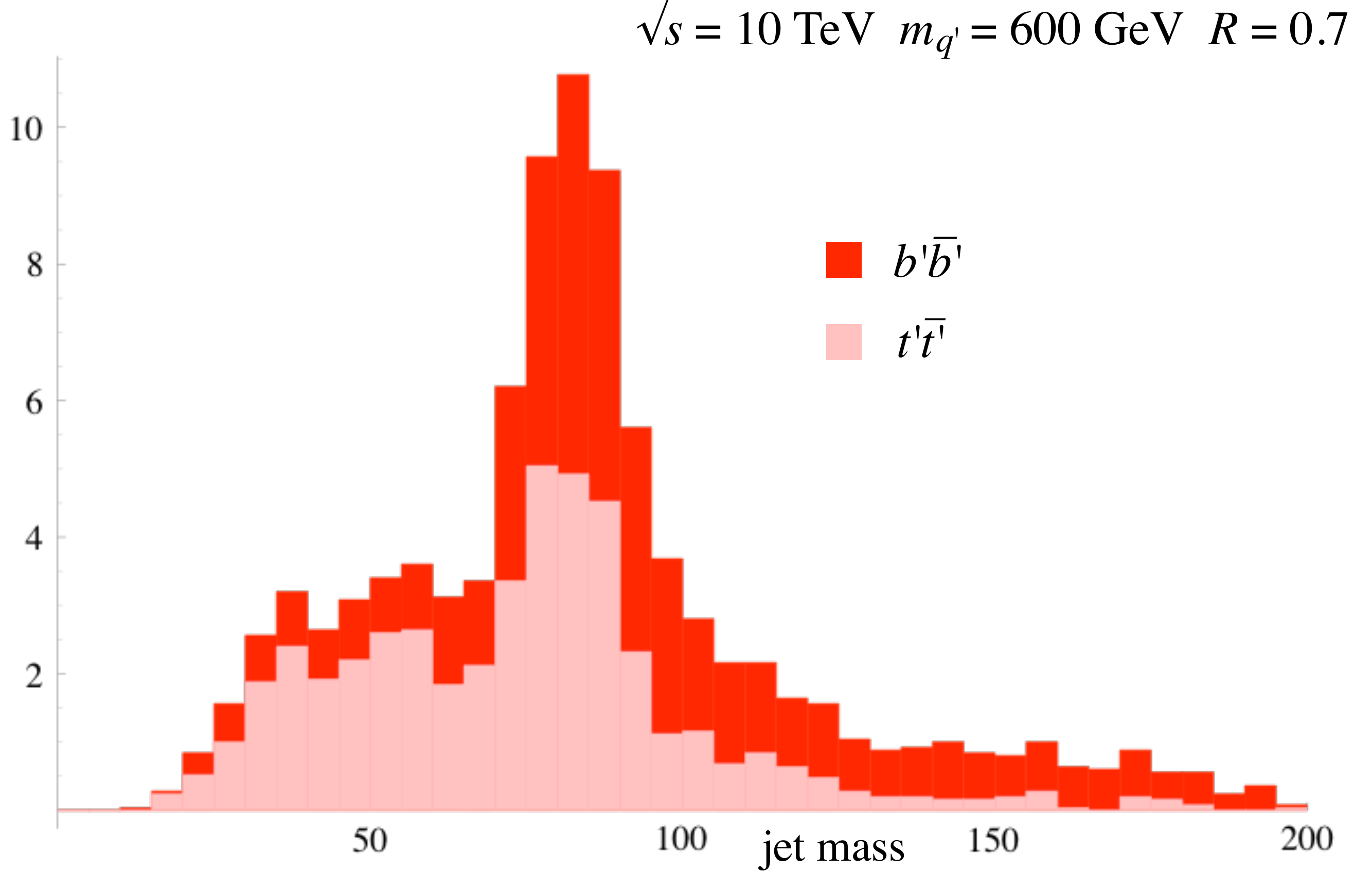}
\end{center}
\vspace{-1ex}\noindent Figure 1: Largest jet mass (signal only)
\vspace{2ex}
\begin{center}\includegraphics[scale=0.4]{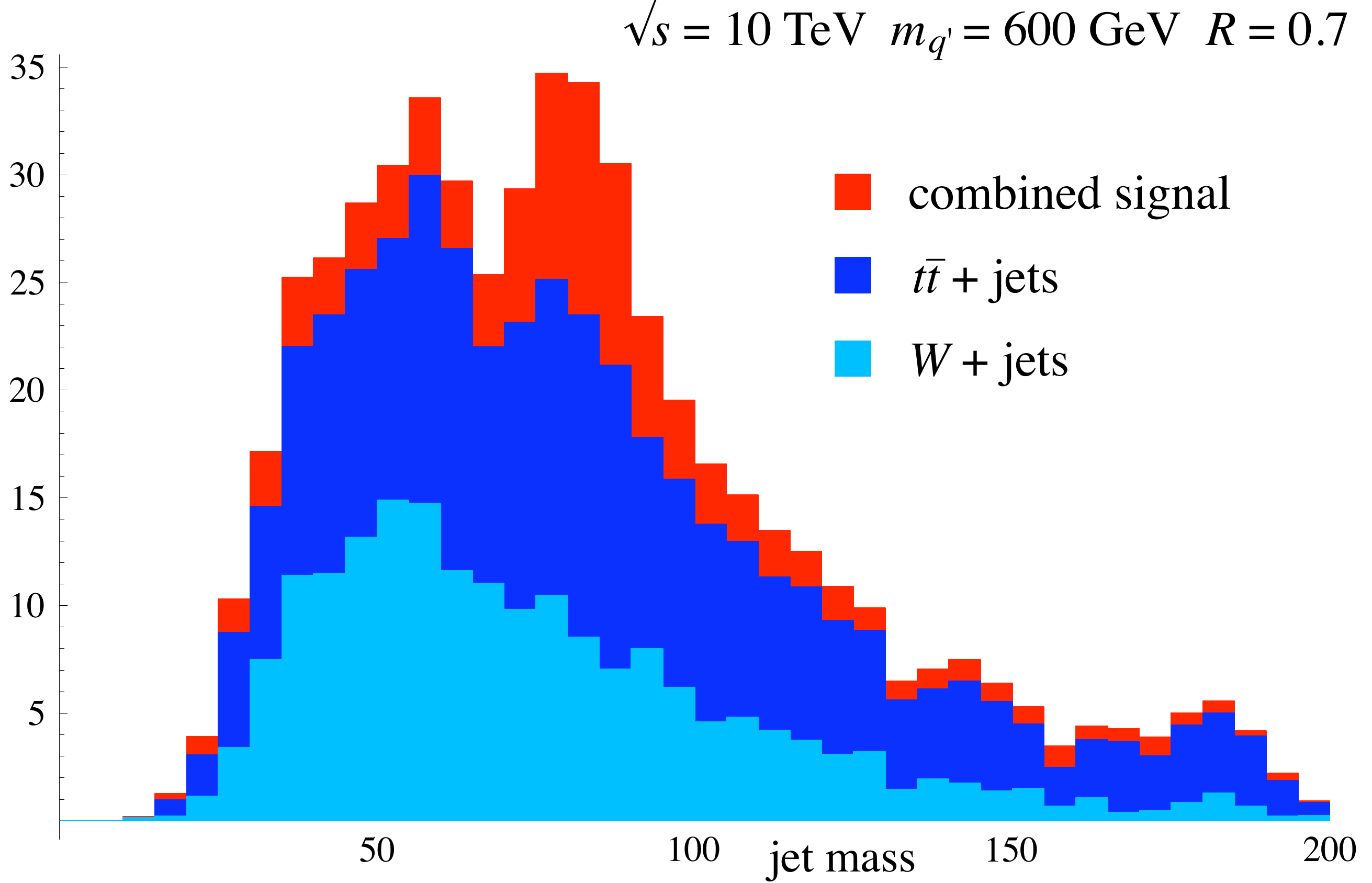}
\end{center}
\vspace{-1ex}\noindent Figure 2: Largest jet mass (combined signal and background)
\vspace{2ex}

We see that a strong $W$ mass peak does not show up in the background events. This is the key for isolating signal from background since it makes possible some type of side-band subtraction. It reflects the fact that this method is only really effective for identifying $W$'s that are both boosted and isolated. There are boosted $W$'s decaying hadronically in the $t\overline{t}+\mbox{}$jets background, but they are typically not isolated due to a neighboring $b$ jet.  For the $W+\mbox{}$jets contribution the $W$ is required to decay leptonically to satisfy event selection, so while QCD jets can fluctuate into large invariant masses this does not produce a $W$ peak. The $W+\mbox{}$jets background could also be significantly reduced when $b$ tagging becomes available.
\begin{center}\includegraphics[scale=0.4]{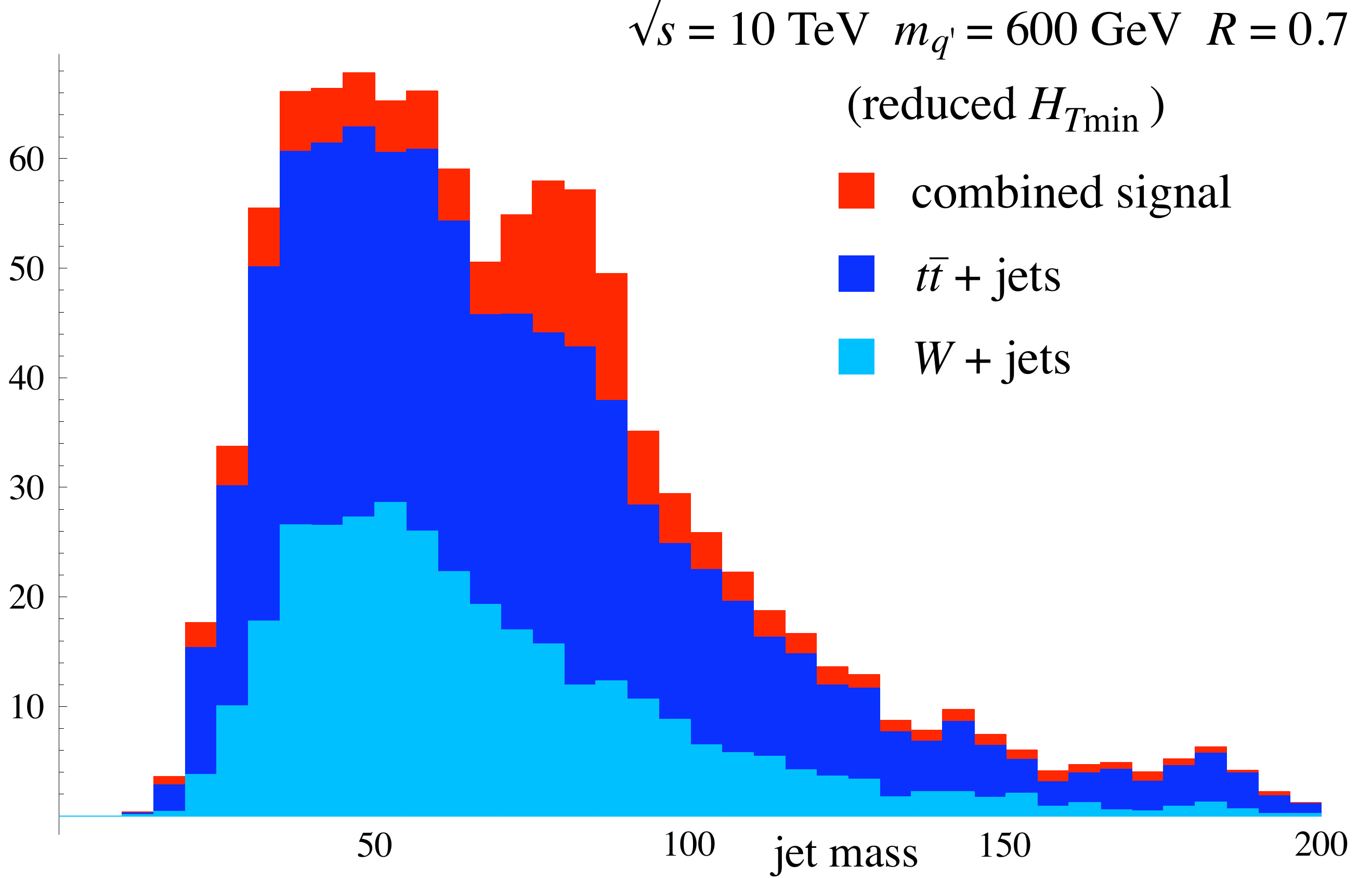}
\end{center}
\vspace{-1ex}\noindent Figure 3: Largest jet mass ($H_{T{\rm min}}$ reduced from 1200 GeV to 1000 GeV)
\vspace{2ex}
\begin{center}\includegraphics[scale=0.4]{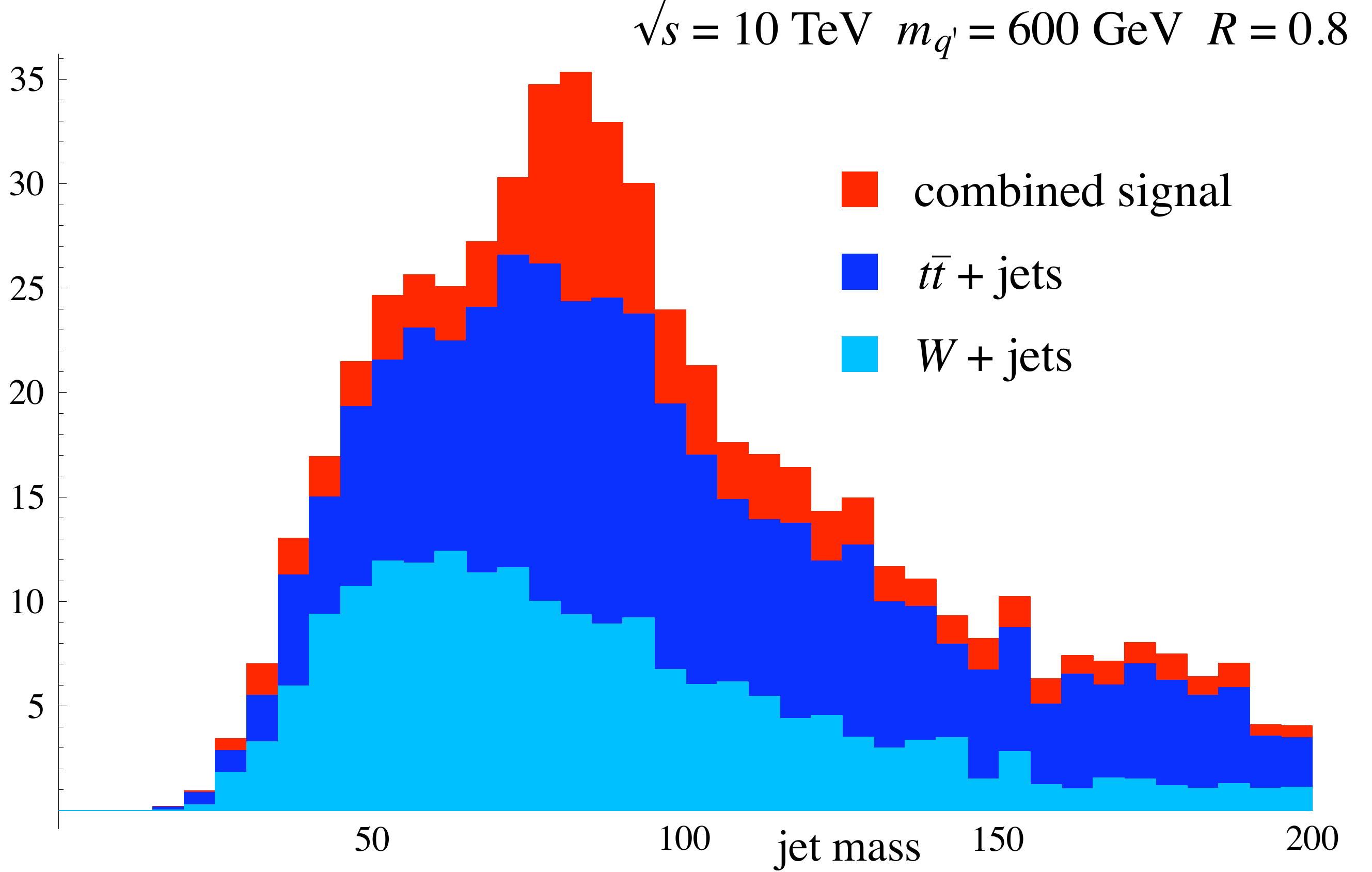}
\end{center}
\vspace{-1ex}\noindent Figure 4: Largest jet mass ($R$ increased)
\vspace{2ex}

We next consider a reduced heavy quark mass of $m_{q'}=450$ GeV, and to place this in the context of the very early running of the LHC we also reduce the center of mass of energy to $\sqrt{s}= 7$ TeV. The cuts are proportionally reduced to $p_T>75$ GeV for jets and $H_{T{\rm min}}=900$ GeV. In Figs.~(5) and (6) we display results for $R=0.8$ and $R=0.9$. These larger $R$ values are now more effective due to the typically smaller boosts of the $W$'s, and this illustrates further the way the shapes of the distributions as a function of $R$ will give information about $m_{q'}$.
\begin{center}\includegraphics[scale=0.4]{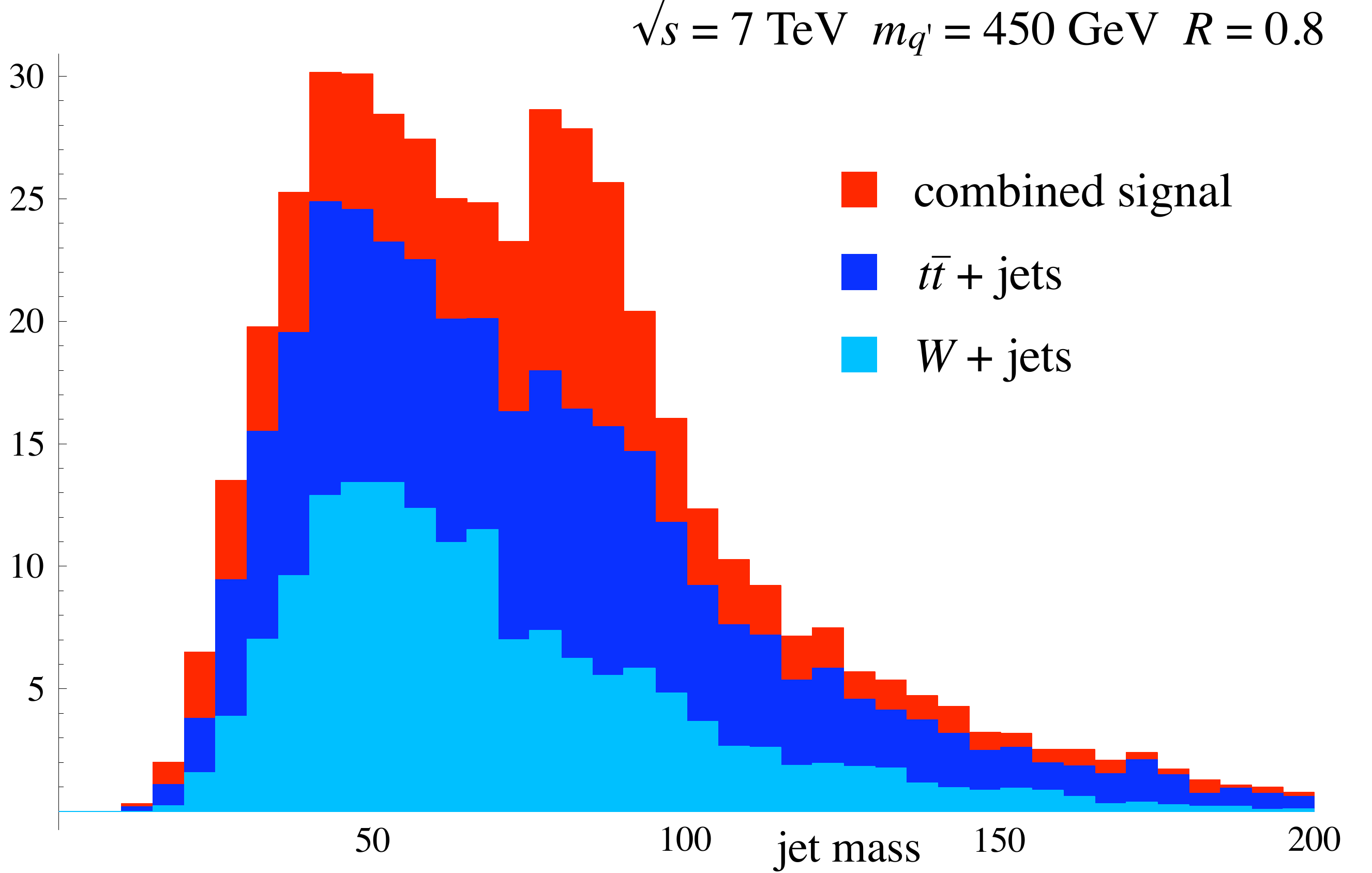}
\end{center}
\vspace{-1ex}\noindent Figure 5: Largest jet mass (smaller $\sqrt{s}$ and $m_{q'}$)
\vspace{2ex}
\begin{center}\includegraphics[scale=0.4]{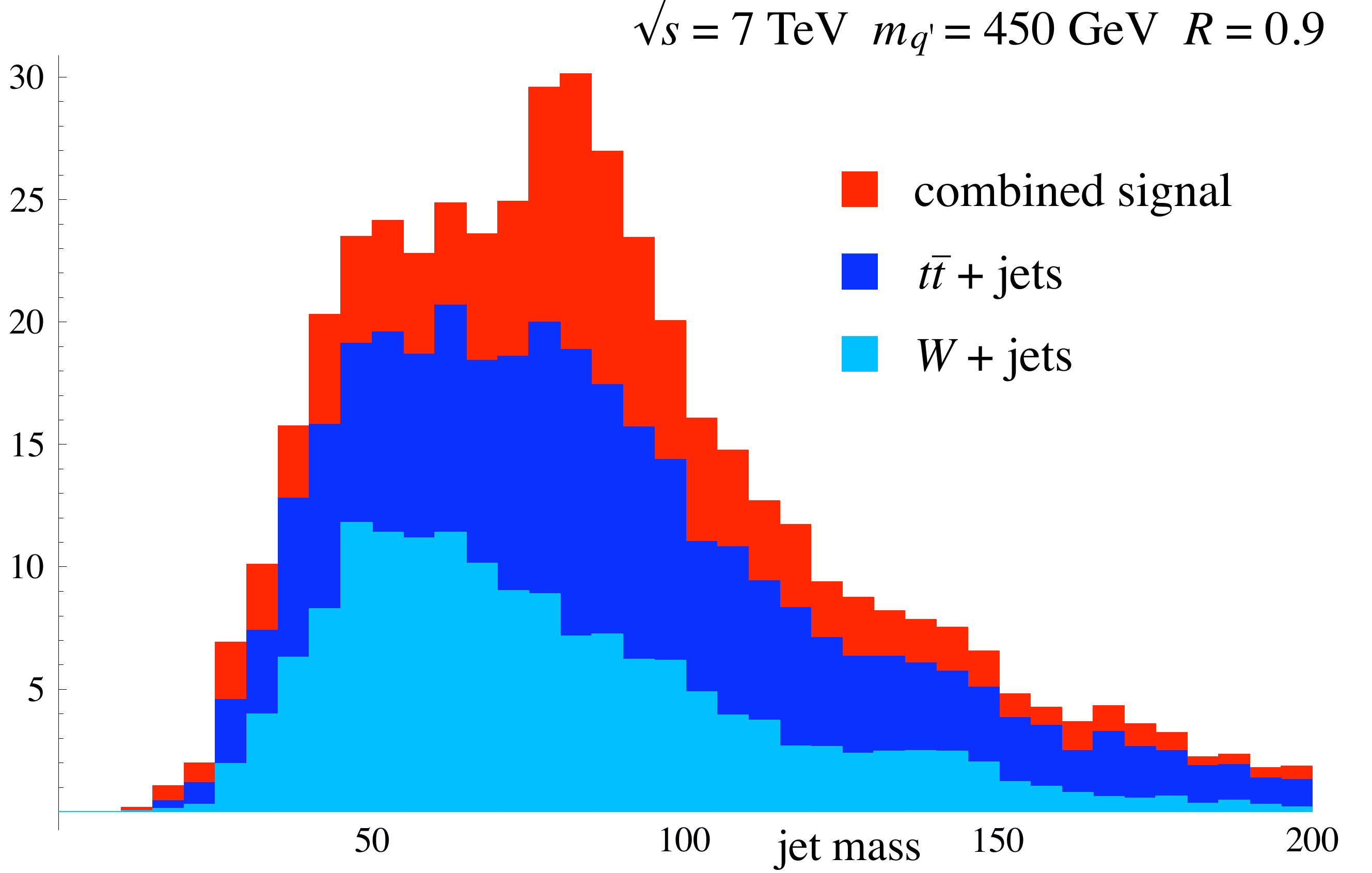}
\end{center}
\vspace{-1ex}\noindent Figure 6: Largest jet mass ($R$ increased)
\vspace{2ex}

Given the success of identifying $W$-jets from the signal events, we see why $W$-jets may be useful for a direct reconstruction of the heavy quark masses \cite{a2,a21,a3}. We can show this here for the present set of parameters, and including both the $t'$ and the $b'$ contributions. The basic idea is to combine a $W$-jet with one other jet. Events are selected to have only three jets having $p_T>150$ GeV and $|\eta|<2.5$, labelled $J_W$, $J_1$, $J_2$. The $W$-jet, $J_W$, has a jet mass within 12 GeV of $M_W$. With $L$ defined as before we require that $L$ be $\Delta R>1$ away from $J_W$, $J_1$, $J_2$. We consider all possible ways of assigning objects to [$J_W$, $J_1$, $J_2$, $L$]. In each case we let the invariant mass of [$J_W$, $J_1$] contribute to the histogram. Note that for the events which contribute, they each contribute at least twice to the histogram. This technique helps to avoid biasing the background contributions, thus suppressing a false peak in the background that often arises in traditional attempts at full event reconstruction.
\begin{center}\includegraphics[scale=0.4]{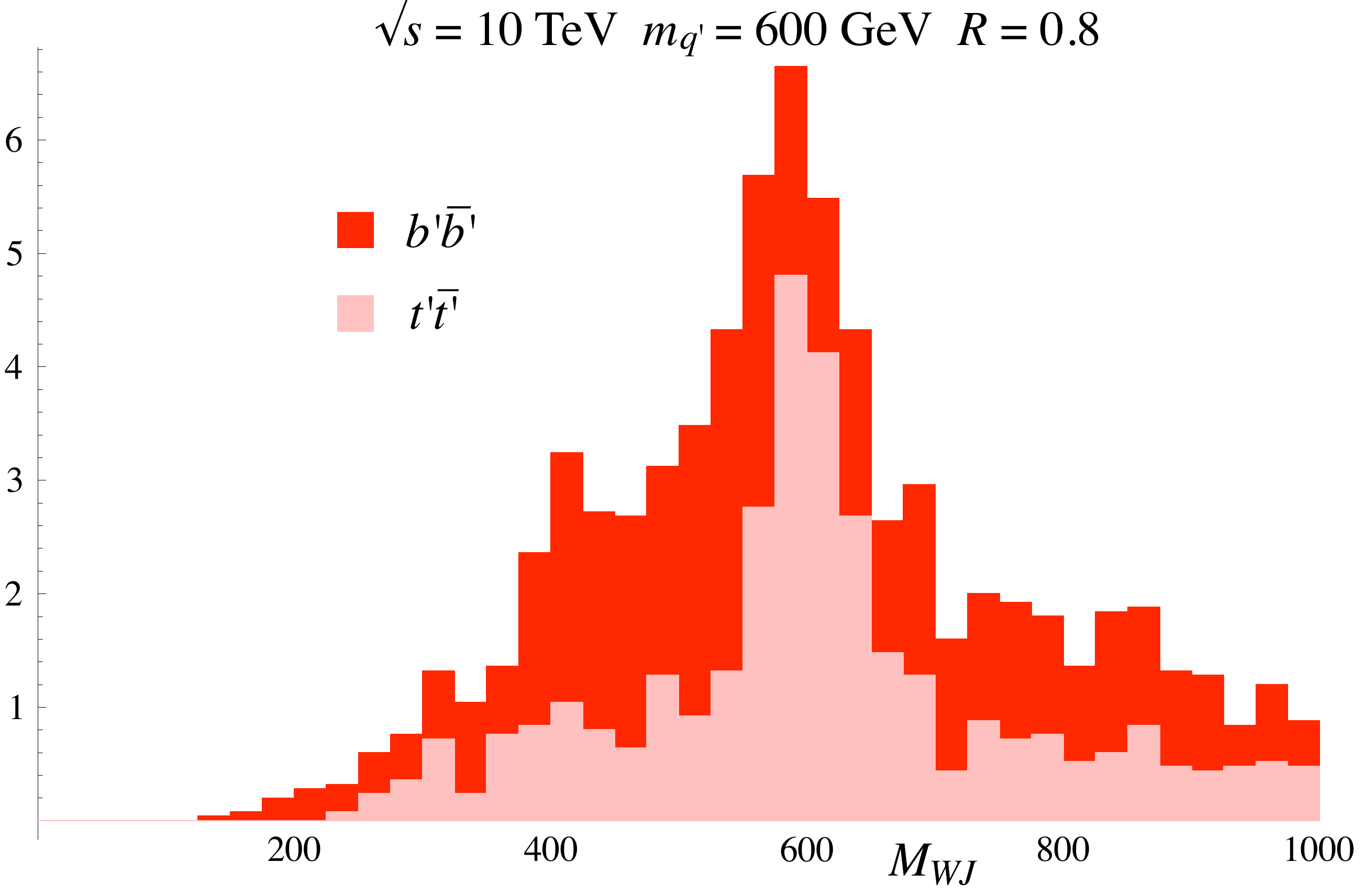}
\end{center}
\vspace{-1ex}\noindent Figure 7: Invariant mass of $W$-jet $+$ jet (signal only)
\vspace{2ex}
\begin{center}\includegraphics[scale=0.4]{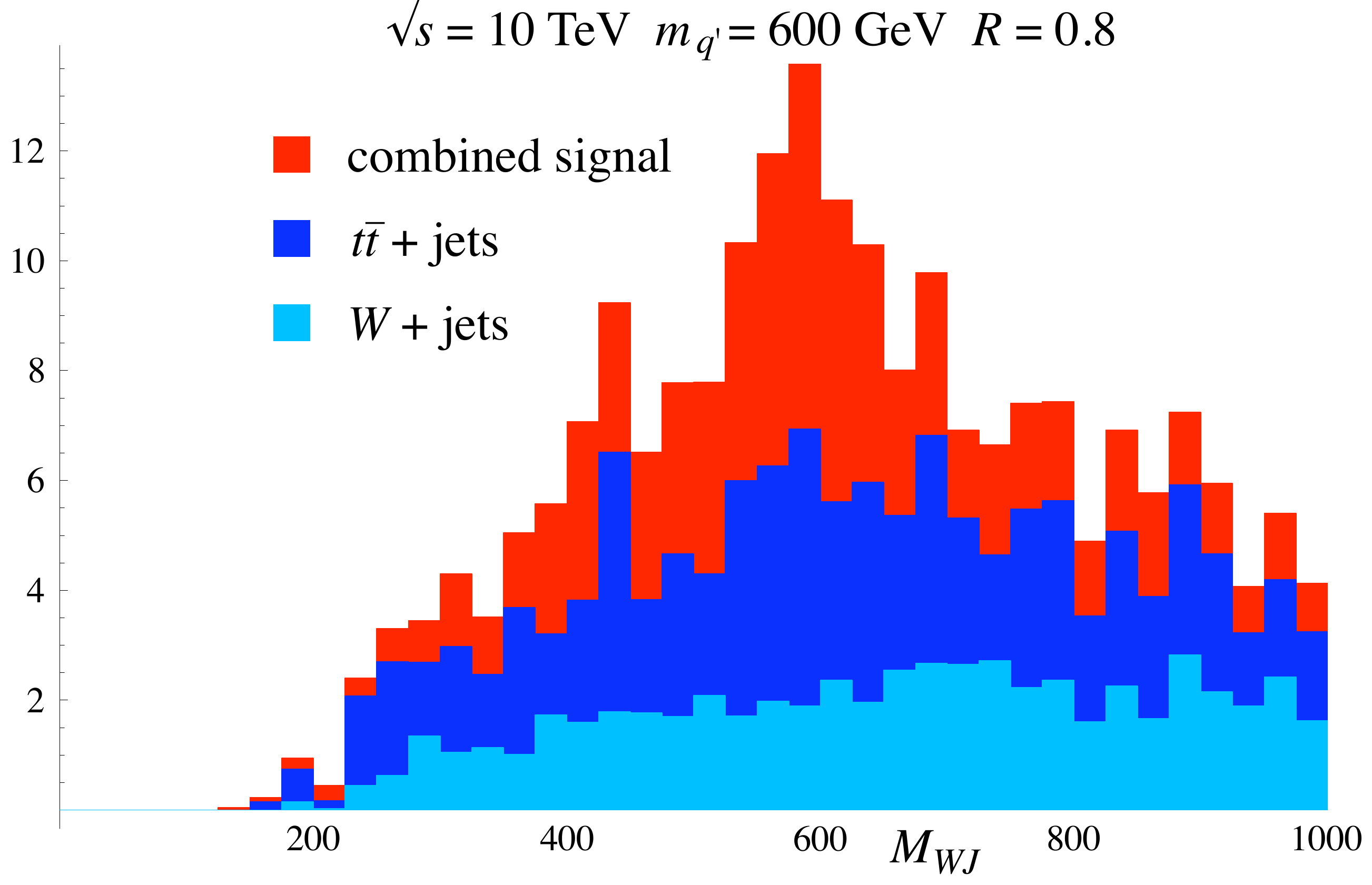}
\end{center}
\vspace{-1ex}\noindent Figure 8: Invariant mass of $W$-jet $+$ jet (combined signal and background)
\vspace{2ex}

Fig.~(7) shows the peak in the reconstructed heavy quark mass for the signal events. The $t'\overline{t'}$ contribution gives a more realistic peak while the $b'\overline{b'}$ contribution is broader and extends to invariant masses below $m_{q'}$. This is understandable given that the $J_1$ jet for the latter case is attempting to capture the hadronic decay of the $t$, where the $t$ is often not boosted enough for this to be effective. It is of interest to contrast the $t'\overline{t'}$ dominance in the heavy quark mass peak to the roughly equal $t'\overline{t'}$ and $b'\overline{b'}$ contributions in the previous $W$ mass peak. The relative strength of the $W$ and $q'$ mass peaks could then in principle be used as evidence for the existence of both the $t'$ and the $b'$. Fig.~(8) shows the combined signal and background while Fig.~(9) shows the same for different choices of parameters.
\begin{center}\includegraphics[scale=0.4]{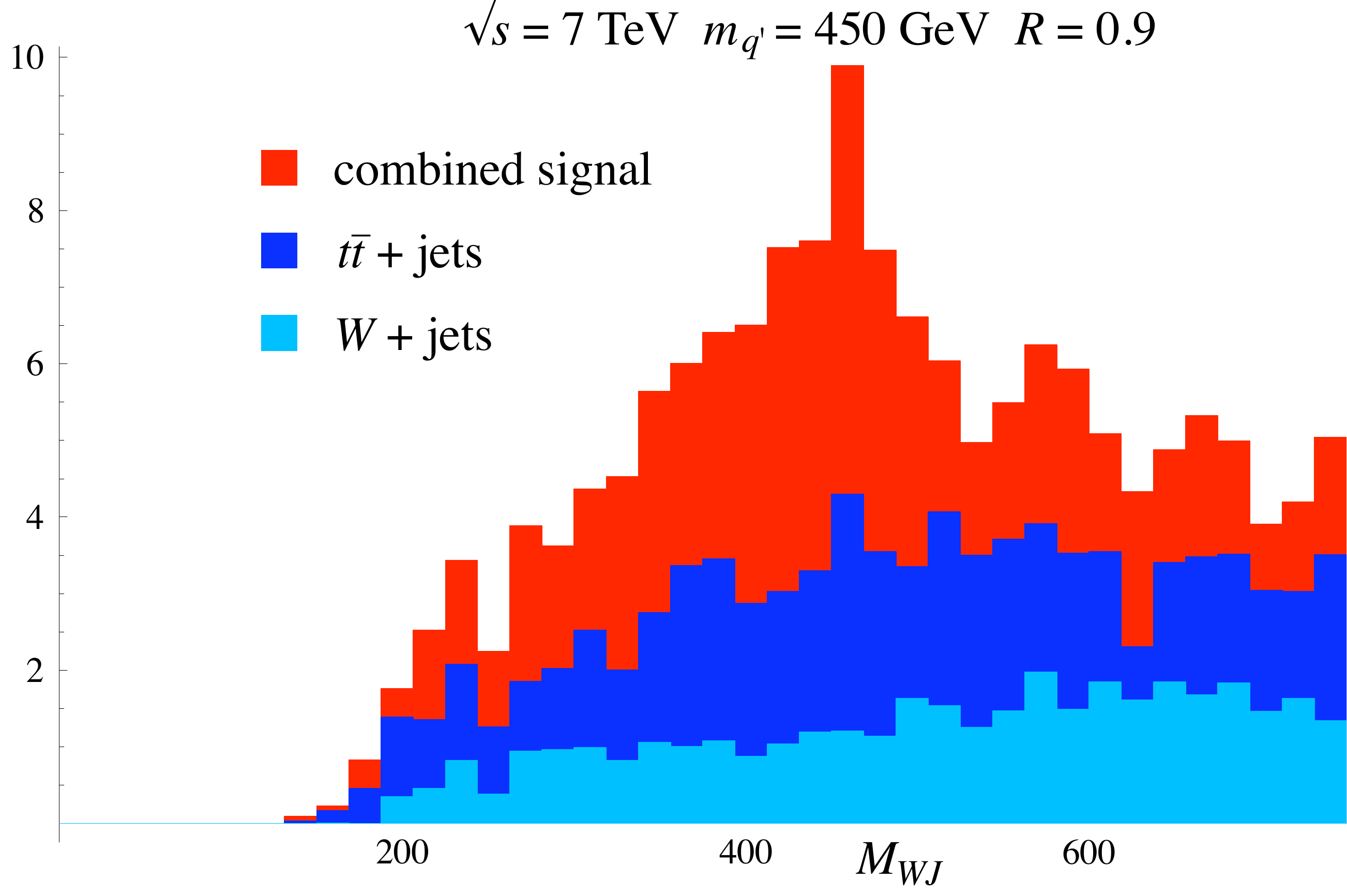}
\end{center}
\vspace{-1ex}\noindent Figure 9: Invariant mass of $W$-jet $+$ jet (smaller $\sqrt{s}$ and $m_{q'}$ and larger $R$)
\vspace{2ex}

We comment here on our use of large jet resolution parameters, $R$ in the range of 0.7 to 0.9, and the possibility that this may imply sensitivity to multiple parton interactions and the underlying event. To test this we changed the underlying event model in Pythia in the generation of the signal events. We changed from tune 108 (old model) to tune 326 (new model) and we found that this hardly degrades the signal. We also turned off the multiple parton interactions altogether, in which case the $W$ and heavy quark mass peaks are only somewhat sharper.

We have described methods that will allow a fairly early discovery of heavy quarks with standard charged current decays. The next question is how to distinguish such sequential quarks from vector quarks. To illustrate the issues we consider a vector doublet of quarks $Q=(T,B)$ because this most closely resembles the sequential quarks $q'=(t',b')$. The mixing of the vector quarks with $t$ and $b$ happens through Yukawa couplings
\begin{equation}
{\cal L}_{\rm mixing}=\textcolor{red}{Y_t}\overline{Q}_Lt_R\tilde{\phi}+\textcolor{blue}{Y_b}\overline{Q}_Lb_R\phi+hc.
\end{equation}
Since the Goldstone bosons are identified with the longitudinal modes of the gauge bosons, this mixing implies the following decay modes.
\begin{eqnarray}
&&T\rightarrow\textcolor{blue}{W^+b,\;}\textcolor{red}{Zt, Ht}\nonumber\\
&&B\rightarrow\textcolor{red}{W^-t}\textcolor{blue}{,Zb, Hb}
\end{eqnarray}
As indicated by the color coding, $\color{red}Y_t$ ($\color{blue}Y_b$) controls the branching fraction into modes involving $\color{red}t$ ($\color{blue}b$). But regardless of the relative size of $Y_t$ and $Y_b$, and for $T$ and $B$ of similar mass, the mixing as described implies that the proportions of $W$, $Z$, and $H$ produced will be close to $1:1/2:1/2$. This is also true of other varieties of vector quarks and top partners.

This leads us to compare the production of like-sign leptons to the production of pairs of leptons that arise due to $Z$ decay. The former can arise when at least three $W$'s are produced, as happens when at least one of the two heavy quarks decays to $Wt$. Thus like-sign leptons can occur for both sequential and vector quarks, while leptons from $Z$ decay occur only for vector quarks. For suitable event selection the backgrounds for both processes can be made small. We therefore need only count the signal events, and we consider the case $\sqrt{s}=10$ GeV and $m_{q'}=600$ GeV with 1 fb$^{-1}$.  

Event selection for leptons from $Z$ is: $H_T>1$ TeV, [2 isolated leptons and $E\!\!\!\!/>$ 100 GeV] or [3 isolated leptons], and $M(e^\pm e^\mp)$ or $M(\mu^\pm \mu^\mp)$ within 4 GeV of the $Z$ mass. The requirement of missing energy or an extra lepton eliminates the $Z+\mbox{}$jets background. This leaves $WZ+\mbox{}$jets as the largest of the small backgrounds, and this could be reduced further with a $b$-tag. Event selection for like-sign leptons is:  $H_T>1$ TeV,  2 isolated same-sign leptons, $E\!\!\!\!/>$ 50 GeV and $M(\ell^\pm\ell^\pm)>100$ GeV. The backgrounds for this process are considered in \cite{a6,a0}.

We find for the number of events, without a $b$-tag:
\begin{center}\begin{tabular}{|c|c|c|}\hline & $\ell^\pm\ell^\pm$ & $Z\rightarrow\ell^+\ell^-$ \\\hline \mbox{sequential quarks} & 7 & 0.7 \\\hline \mbox{vector quarks} & \mbox{1 to 7} & 6.3 \\\hline $WZ+\mbox{}$jets & 0 & 1.5 \\\hline \end{tabular}\end{center}
The 1 to 7 corresponds to $Y_t/Y_b$ varying from 0 to $\infty$. These results indicate that sequential quarks could be distinguished from vector quarks at roughly the same time that a new same sign lepton signal is detected. Thus when heavy quark masses have been reconstructed or inferred and if leptons from $Z$ are not observed at the rate at least comparable to same sign leptons, then we will have good evidence for new quarks belonging to a fourth family.

\section{Implications}
The existence of a fourth family would cast a new light on the question of neutrino mass. A sufficiently heavy fourth neutrino is required and is at first sight strange; this has even held back the consideration of a fourth family. A large neutrino mass is also often taken to imply that a large Dirac mass is present (with or without a separate Majorana mass for the right-handed neutrino). This bias may also be faulty since there may well be no right handed neutrinos, at least at or below the TeV scale. In particular no known gauge symmetry protects the $\nu_R$'s from receiving some mass much larger than a TeV, from whatever dynamics existing at higher scales.

When we discussed the origin of mass for a heavy fourth family we described a new strong interaction that coupled to all members of the fourth family. Thus the fourth neutrino mass could arise in a way similar to the other fourth family fermions, and in the absence of $\nu_R$'s this would imply a Majorana mass for $\nu'_{L\tau}$. In any case a fourth neutrino mass similar to the other fourth family members is not in itself unnatural, and the real question is why the other neutrinos are so light.

Before pursuing that question we can look at the implications for electroweak precision observables. The effect of the Majorana mass for a purely left-handed neutrino shows up in the last term in each of the following contributions to $S$ and $T$ \cite{a1,a7}.
\begin{eqnarray}
S_{\rm leptons}&\approx&\frac{1}{6\pi}-\frac{1}{3\pi}\ln(\frac{m_{\tau'}}{m_{\nu'}})-\color{red}{\frac{1}{12\pi}}\label{e5}\\
\alpha v^2 T_{\rm leptons}&\approx&\frac{1}{12\pi^2}(m_{\tau'}-m_{\nu'})^2-\color{red}\frac{m_{\nu'}^2}{4\pi^2}\ln(\frac{\Lambda_{\nu'}}{m_{\nu'}})\label{e6}
\end{eqnarray}
The presence of $\Lambda_{\nu'}$ (just somewhat larger than $m_{\nu'}$) is associated with the dynamical nature of the mass and the fact that the mass function falls off quickly above this scale. Fig.~(10) illustrates these leptonic contributions as a function of $m_{\tau'}$ while holding the other fourth family masses fixed.
\begin{center}\includegraphics[scale=0.5]{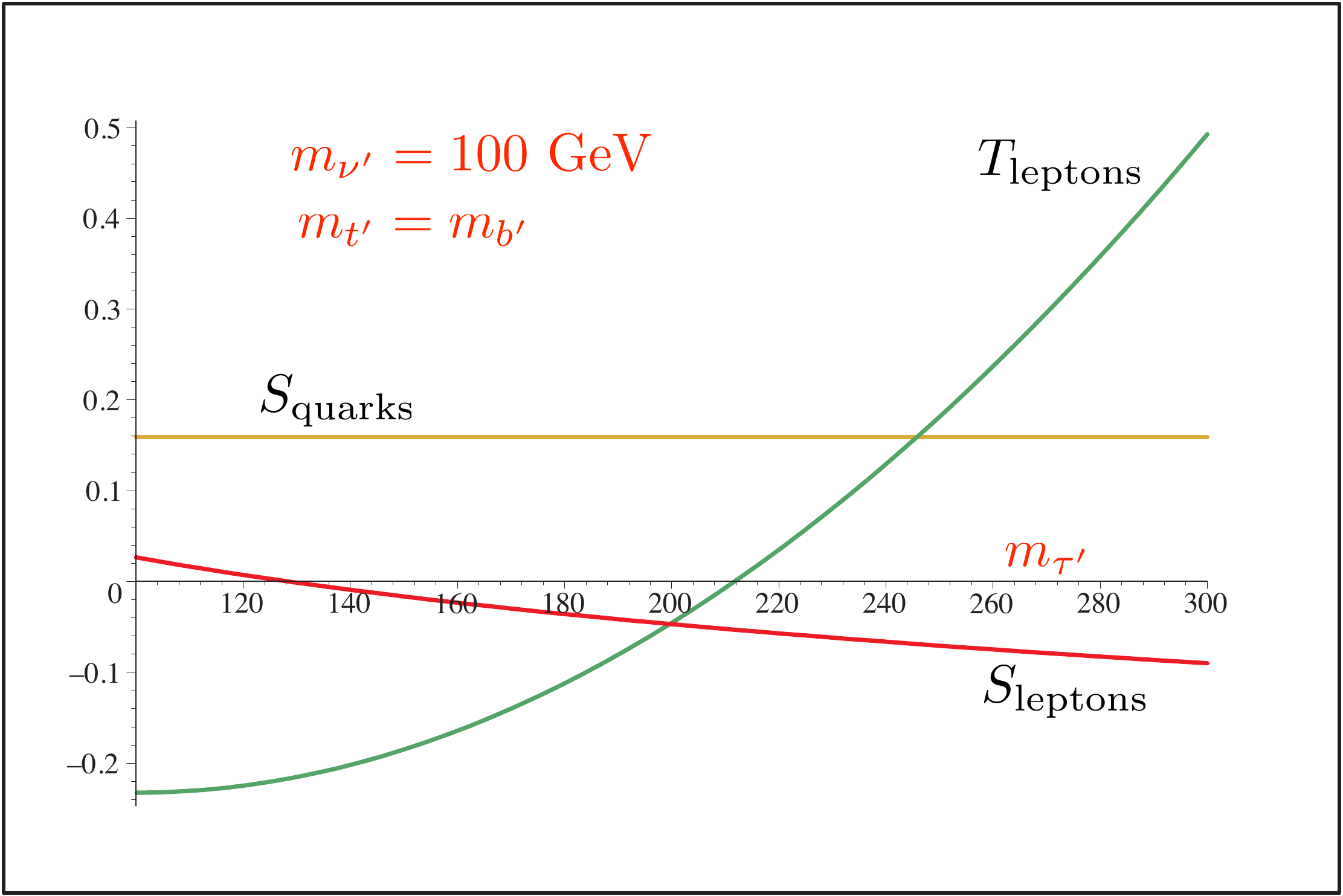}
\end{center}
\vspace{-1ex}\noindent Figure 10: Illustration of equations (\ref{e5}) and (\ref{e6}).
\vspace{2ex}

In the absence of a light Higgs, a contribution to $T$ of the order 0.5 is needed. Since $T_{\rm leptons}$ is decreased by the new term, one needs a larger $m_{\tau'}/m_{\nu'}$ to obtain $T_{\rm leptons}\approx0.5$. One is forced to the right side of the plot which decreases $S_{\rm leptons}$ further as shown, and this can thus cancel a larger portion of $S_{\rm quarks}$. Larger values of $m_{\tau'}$ and $m_{\nu'}$ can also produce this effect, and the result is that any constraint on the relative size of $m_{t'}$ and $m_{b'}$ is lessened. Thus while there was never a serious conflict between a fourth family  and electroweak precision observables \cite{a6a,a6b,a1,a7,a2a}, a $\nu'_{L\tau}$ with a Majorana mass makes a fourth family even easier to accommodate \cite{a1,a7}.

Let us return to the question of light neutrino masses by comparing them to their associated charged lepton masses. We relate the much lighter neutrino masses to our hypothesis that they are purely left handed. Consider the following operators (which can be written in an $SU(2)_L\times U(1)_Y$ symmetric form) that feed down mass from the $\tau'$ to a light charged lepton $e$ (or $\mu$) and a light neutrino $\nu_e$ (or $\nu_\mu$).
\begin{equation}
\frac{1}{\Lambda^2_{fl}}\overline{\tau'}_L\tau'_R\overline{e}_Re_L
\label{e1}\end{equation}
\begin{equation}
\frac{1}{\Lambda^5_{fl}}\overline{\tau'}_R\tau'_L\overline{\tau'}_R\tau'_L\nu_{Le}\nu_{Le}
\label{e2}\end{equation}
The main point is that there is no 4-fermion operator that can feed mass from $\tau'$ (or $t'$ or $b'$) to the $\nu_e$. Then for example if $\Lambda_{fl}\approx200$ TeV and $\langle\overline{\tau'}_L \tau'_R\rangle\approx(\mbox{300 GeV})^3$, the resulting illustrative lepton masses are $m_e\approx\mbox{0.7 MeV and }m_\nu\approx0.002$ eV.

We note that the 6-fermion operator could result from 4-fermion operators that involve right-handed neutrinos after the latter are integrated out. Thus it is quite consistent for the $\nu_R$ masses to be similar to the highest flavor scale, denoted here by $\Lambda_{fl}$.\footnote{We have ignored possible anomalous scaling enhancement of the operators in (\ref{e1}) and (\ref{e2}), which if present would increase $\Lambda_{fl}$.} The large Majorana mass for right handed neutrinos is a natural order parameter for the breaking of the original flavor gauge symmetries, including a possible $SU(2)_R$. We can contrast this situation to the standard see-saw with Higgs, which requires a much larger hierarchy between the $\nu_R$ masses and other masses (or alternatively some very small Yukawa couplings).

There is in fact a 4-fermion operator that could feed down mass from the $\nu'_{L\tau}$ to the light neutrinos, of the form $\overline{\nu'}_{L\tau}\overline{\nu'}_{L\tau}\nu_L\nu_L$. This can also be written in an $SU(2)_L\times U(1)_Y$ symmetric form and the $\nu_L$ here can be the $e$, $\mu$ or $\tau$ neutrino. But this operator can be eliminated, or strongly suppressed, if (fourth family lepton number) $-$ (third family lepton number) is a symmetry, or a very good approximate symmetry of the flavor physics. The operators in (\ref{e1}) and (\ref{e2}) are allowed by this symmetry, while it is broken by the $\nu'_{L\tau}$ dynamical mass. This symmetry may correspond to a broken gauge symmetry, in which case the associated gauge boson may be a TeV scale remnant of the original flavor gauge symmetry.

Similarly we consider 4-fermion operators that can feed down mass from the fourth family quarks to the third family quarks.
\begin{equation}
\overline{t'}_Lt'_R\overline{t}_Rt_L\quad\quad\overline{b'}_Lb'_R\overline{b}_Rb_L
\label{e3}\end{equation}
\begin{equation}
\overline{b'}_Lb'_R\overline{t}_Lt_R\quad\quad\overline{t'}_Lt'_R\overline{b}_Lb_R
\label{e4}\end{equation}
Here we note the effect of an approximate symmetry defined by the charges:
$$\begin{array}{|c|c|c|c|}\hline  t'_L  & t'_R & b'_L &b'_R \\\hline + & - & + & - \\\hline t_L  & t_R & b_L &b_R\\\hline - & + & - & +\\\hline \end{array}$$
This is (fourth family axial quark number) $-$ (third family axial quark number) and it may be the quark part of the same remnant flavor gauge symmetry already mentioned. The $q'$ masses also contribute to its breaking. To the extent that it is a symmetry of the flavor physics it would suppress the two operators in (\ref{e3}) relative to the two operators in (\ref{e4}). This is advantageous since the $t$ mass operator in (\ref{e4}) has suppressed impact on the $Zb\overline{b}$ coupling; two insertions of this operator are needed due to its LRLR form \cite{a8}. We mention in passing that the massive gauge boson associated with the remnant flavor gauge symmetry has its own implications for LHC physics \cite{a9}.

We have hopefully made it clear why a fourth family is interesting, and especially so if its masses are large enough to imply new strong interactions. Despite the presence of strong interactions, the standard weak decays of the heavy quarks make them quite easy to find or to rule out at the LHC. Any modification of the production cross section due to the strong interactions should not change this conclusion. A crucial step will then be to distinguish fourth family quarks from vector quarks, for example in a manner we have described. The confirmation of heavy sequential quarks would bring an end to the long ``Era of the Higgs''. It would cause the focus of the LHC, which has been to find the Higgs, to suddenly shift. And it would also cause the focus of theorists to suddenly shift. Theorists would see that new strong dynamics is showing up at the TeV scale, right where it is needed. This would be a serious blow to anthropic thinking, and the concept of naturalness may return as an important criteria in the construction of physical theories.

\section*{Acknowledgments}
I thank Wei-Shu Hou and the other organizers and participants for a fruitful workshop experience. I also thank Qi-Shu Yan for numerous discussions. This work was supported in part by the Natural Science and Engineering Research Council of Canada.

\end{document}